\documentclass[12pt,a4paper]{article}
\usepackage{amsmath,amssymb,psfrag,graphicx}
\usepackage{hyperref}
\begin{document}
\input{epsf}
\topmargin 0pt
\oddsidemargin 5mm
\headheight 0pt
\headsep 0pt
\topskip 9mm
\pagestyle{empty}

\newcommand{\beq}{\begin{equation}}
\newcommand{\eeq}{\end{equation}}
\newcommand{\bea}{\begin{eqnarray}}
\newcommand{\eea}{\end{eqnarray}}
\newcommand{\rf}[1]{(\ref{#1})}
\newcommand{\pa}{\partial}
\newcommand{\nn}{\nonumber}
\newcommand{\e}{\mbox{e}}
\renewcommand{\d}{\mbox{d}}
\newcommand{\g}{\gamma}
\renewcommand{\l}{\lambda}
\renewcommand{\L}{\Lambda}
\renewcommand{\b}{\beta}
\renewcommand{\a}{\alpha}
\newcommand{\n}{\nu}
\newcommand{\m}{\mu}
\newcommand{\Tr}{\mbox{Tr}}
\newcommand{\E}{\mbox{E(q)}}
\newcommand{\Ee}{\mbox{E}}
\newcommand{\K}{\mbox{K(q)}}
\newcommand{\Kk}{\mbox{K}}
\newcommand{\ep}{\varepsilon}
\newcommand{\om}{\omega}
\newcommand{\del}{\delta}
\newcommand{\Del}{\Delta}
\newcommand{\sg}{\sigma}
\newcommand{\vph}{\varphi}
\newcommand{\sn}{\mbox{sn}}
\newcommand{\dn}{\mbox{dn}}
\newcommand{\cn}{\mbox{cn}}
\newcommand{\CA}{\mathcal{A}}
\newcommand{\oh}{\frac{1}{2}}
\newcommand{\oq}{\frac{1}{4}}
\newcommand{\dg}{\dagger}
\newcommand{\R}{\mathbb{R}}
\newcommand{\tr}{\mbox{Tr}\;}
\newcommand{\ra}{\right\rangle}
\newcommand{\la}{\left\langle}
\newcommand{\prt}{\partial}
\newcommand{\mi}{\!-\!}
\newcommand{\equ}{\!=\!}
\newcommand{\pl}{\!+\!}
\newcommand{\CN}{\mathcal{N}}
\newcommand{\cD}{{\cal D}}
\newcommand{\cS}{{\cal S}}
\newcommand{\cM}{{\cal M}}
\newcommand{\cK}{{\cal K}}
\newcommand{\cT}{{\cal T}}
\newcommand{\cN}{{\cal N}}
\newcommand{\cL}{{\cal L}}
\newcommand{\cO}{{\cal O}}
\newcommand{\cR}{{\cal R}}
\newcommand{\CH}{\mathcal{H}}
\newcommand{\CL}{\mathcal{L}}
\newcommand{\CO}{\mathcal{O}}
\newcommand{\CI}{\mathcal{I}}
\newcommand{\CT}{\mathcal{T}}
\newcommand{\CS}{\mathcal{S}}
\newcommand{\CM}{\mathcal{M}}
\newcommand{\CQ}{\mathcal{Q}}
\newcommand{\CE}{\mathcal{E}}
\newcommand{\CB}{\mathcal{B}}
\newcommand{\tF}{{\tilde{F}}}
\newcommand{\tL}{{\tilde{\L}}}
\newcommand{\tX}{{\tilde{X}}}
\newcommand{\tY}{{\tilde{Y}}}
\newcommand{\tZ}{{\tilde{Z}}}
\newcommand{\ty}{{\tilde{y}}}
\newcommand{\tz}{{\tilde{z}}}
\newcommand{\tg}{{\tilde{g}}}
\newcommand{\tG}{{\tilde{G}}}
\newcommand{\tH}{{\tilde{H}}}
\newcommand{\tT}{{\tilde{T}}}

\newcommand{\SL}{{\sqrt{\L}}}
\newcommand{\tSL}{\sqrt{\tL}}
\newcommand{\FL}{\L^{1/4}}
\newcommand{\bZ}{{\bar{Z}}}
\newcommand{\bX}{{\bar{X}}}

\newcommand{\remark}[1]{{\renewcommand{\bfdefault}{b}\textbf{\mathversion{bold}#1}}}

\vspace*{100pt}

\begin{center}

{\large \bf {Non-planar ABJ Theory and Parity}}

\vspace*{26pt}

{\sl Pawel Caputa}, {\sl Charlotte Kristjansen} and
{\sl Konstantinos Zoubos}

\vspace{10pt}
\vspace{10pt}

The Niels Bohr Institute, Copenhagen University\\
Blegdamsvej 17, DK-2100 Copenhagen \O , Denmark.\\\vspace{.4cm}
{\tt\small caputa@nbi.dk, kristjan@nbi.dk, kzoubos@nbi.dk}
\vspace{10pt}

\end{center}

\begin{abstract}
\noindent
 While the ABJ Chern--Simons--matter theory and its string theory dual
 manifestly lack parity invariance, no sign of parity violation has so
 far been observed on the weak coupling spin chain side. In particular, the
 planar two-loop dilatation generator of ABJ theory is parity invariant. In 
 this  letter we derive the non-planar part of the two-loop dilatation
 generator of ABJ theory in its $SU(2)\times SU(2)$ sub-sector. Applying
 the dilatation generator to short operators, we explicitly demonstrate
 that, for operators carrying excitations on both spin chains, the
 non-planar part breaks parity invariance.
 For operators with only one type of excitation, however, parity remains
 conserved at the non-planar level.
 We furthermore observe that, as for ABJM theory, the degeneracy between
 planar parity pairs is lifted when non-planar corrections are taken
 into account.

\end{abstract}

\newpage

\pagestyle{plain}

\setcounter{page}{1}

\newcommand{\ft}[2]{{\textstyle\frac{#1}{#2}}}
\newcommand{\ii}{\mathrm{i}}
\newcommand{\dd}{{\mathrm{d}}}
\newcommand{\nnb}{\nonumber}

\section{Introduction}
The concept of spin chain parity~\cite{Doikou:1998jh}
 played a crucial role in the discovery of
higher loop integrability of the planar spectral problem
of ${\cal N}=4$ SYM~\cite{Beisert:2003tq}. For a spin chain state the 
parity operation simply inverts the order of spins at the sites of the chain.
In the field theory language the operation correspondingly inverts the
order of fields inside a single trace operator or equivalently complex
conjugates the gauge group generators.
 ${\cal N}=4$ SYM theory
is parity invariant. In particular, the theory's dilatation generator commutes
with parity. Integrability of the planar spectral problem at one loop order,
discovered first in~\cite{Minahan:2002ve}, implies the existence of a tower
of higher conserved charges. The first of these, while commuting with the
dilatation generator, anti-commutes with parity. As a consequence one finds 
in the planar spectrum pairs of operators with opposite parity but the same
conformal dimension, denoted as planar parity pairs. The fact that these
planar parity pairs survived higher loop corrections constituted the seed
for the unveiling of higher loop 
integrability~\cite{Beisert:2003tq,Beisert:2004hm}. 
When non-planar corrections
were taken into account, parity was still a good quantum number but the
degeneracies between planar parity pairs disappeared~\cite{Beisert:2003tq}.
While not disproving integrability this shows that the standard
construction of conserved charges does not work any more.

The discovery of a novel $AdS_4/CFT_3$ 
correspondence~\cite{Aharony:2008ug,Aharony:2008gk}
has provided us with the  possibility of studying the effects of
parity violation in a 
supersymmetric gauge theory and its dual string theory. A supersymmetric
${\cal N}=6$ Chern--Simons--matter theory with gauge group
$SU(M)_k\times \overline{SU(N)}_{-k}$, where $k$ denotes the Chern--Simons
level, has been found to be dual to
type IIA string theory on $AdS_4\times CP^3$ with a background 
NS $B$--field $B_2$ having non-trivial holonomy on $CP^1\subset CP^3$.
More precisely\footnote{Here we have assumed that $M\geq N$. Quantum consistency of the theory 
requires in addition that $M-N\leq k$ \cite{Aharony:2008gk}.}
\beq
\frac{1}{2\pi}\int_{CP^1\subset CP^3} B_2 =\frac{M-N}{k}.
\eeq
This $B$--field holonomy causes breaking of world-sheet parity for 
$M\neq N$ and results in a string background which breaks target-space parity~\cite{Aharony:2008gk}.  
Correspondingly, the dual field theory does not 
respect three-dimensional parity invariance. For $M=N$ the Chern--Simons--matter theory
is known as ABJM theory whereas the general version is denoted as ABJ theory.
Our aim is to investigate how the parity breaking on the field theory side manifests 
itself in the spin chain language. The first steps in this direction were 
taken in~\cite{Bak:2008vd,Minahan:2009te} where the two-loop planar
dilatation generator of ABJ theory was derived, respectively in an
$SU(4)$ sub-sector and for the full set of fields. However, rather surprisingly, in 
these studies no effects of parity violation were seen. In fact the planar
two-loop dilatation generator of ABJ theory differs from that of
ABJM theory~\cite{Minahan:2008hf,Bak:2008cp,Zwiebel:2009vb} only by an
overall pre-factor. This raises the question of whether the parity symmetry of the spin chain 
has a deeper significance, or is simply an accidental symmetry of the two-loop planar 
approximation. In the present letter we will derive the two-loop non-planar dilatation generator 
of ABJ theory in a $SU(2)\times SU(2)\subset SU(4)$ sub-sector and explicitly
demonstrate parity-breaking effects.

We start by, in section~\ref{summary}, briefly describing ABJ theory and 
subsequently proceed to derive its full (planar plus non-planar) two-loop dilatation 
generator in the $SU(2)\times SU(2)$ sector ~in section~\ref{derivation}. 
As the derivation follows closely that of ABJM theory~\cite{Kristjansen:2008ib} 
we shall be very brief. In 
section~\ref{Results} we explicitly apply the dilatation generator 
to a series of short operators and determine their spectrum. 
In particular, we show
that the non-planar part of the dilatation generator does {\it not} conserve
parity. In addition, we observe a lifting of all planar degeneracies. Finally,
section~\ref{conclusion} contains our conclusion.

\section{ABJ theory \label{summary}}
Our notation will follow that of references~\cite{Benna:2008zy,Bak:2008cp}.
ABJ theory \cite{Aharony:2008gk} (see also \cite{Hosomichi:2008jb} for a discussion at the classical level) 
is a three-dimensional ${\cal N}=6$ superconformal Chern--Simons--matter
theory with gauge group $U(M)_k\times \overline{U(N)}_{-k}$ and $R$-symmetry group
$SU(4)$. The parameter $k$ denotes the Chern--Simons level.
The fields of ABJ theory consist of gauge fields $A_m$ and $\bar{A}_m$,
complex scalars $Y^I$ and Majorana spinors $\Psi_I$, $I\in \{1,\ldots 4\}$.
The two gauge fields $A_m$ and $\bar{A}_m$ 
belong to the adjoint representation of $U(M)$ and $\overline{U(N)}$ respectively.
For $N=M$, ABJ theory reduces to ABJM theory.
The scalars $Y^I$ and the
spinors $\Psi_I$ are bi-fundamental and transform in the
$M\times \overline{N}$ representation of the gauge group and in the fundamental
and anti-fundamental representation of $SU(4)$ respectively. For our
purposes it proves convenient to write the scalars and spinors explicitly
in terms of their $SU(2)$ component fields, i.e.~\cite{Benna:2008zy}
\bea
Y^I &= &\{Z^A,W^{\dg A}\},
\hspace{0.7cm} Y_I^\dg=\{Z_A^\dg,W_A\},\nonumber \\
\Psi_I&=& \{\epsilon_{AB}\,\xi^B\, e^{i\pi/4},
\epsilon_{AB}\,\omega^{\dg B}\, e^{-i\pi/4},\}, \nonumber\\
\Psi^{I \dg}& = &\{-\epsilon^{AB}\,\xi_B^\dg\, e^{-i\pi/4},
-\epsilon^{AB}\,\omega_B\, e^{i\pi/4}
\}, \nonumber
\eea
 where now $A,B\in \{1,2\}$.
Expressed in terms of these fields the action reads
\bea
S &= &\int d^3x \left [\frac{k}{4\pi} \epsilon^{m n p} \Tr (
A_m \partial_n A_p+\frac{2i}{3} A_m A_n A_p  )-
\frac{k}{4\pi} \epsilon^{m n p} \Tr (
\bar{A}_m \partial_n \bar{A}_p+\frac{2i}{3} \bar{A}_m \bar{A}_n
\bar{A}_p  ) \nonumber
\right. \\
&& \left.
- \Tr ( {\cal D}_m Z)^\dg {\cal D}^m Z-\Tr ({\cal D}_m W)^\dg {\cal D}^m W
+i \Tr \xi^\dg {\cal D}\hspace{-0.3cm}\slash
\hspace{0.13cm}
\xi +i\Tr \omega^\dg {\cal D}\hspace{-0.3cm}\slash
\hspace{0.13cm}\omega  \nonumber \right.\\
&& \left.
-V_D^{ferm}-V_D^{bos}
-V_F^{ferm}-V_F^{bos}\right]. \nonumber
\eea
Here the covariant derivatives are defined as
\beq
{\cal D}_m Z^A = \partial_m  Z^A + i A_m Z^A-i Z^A \bar{A}_m,
\hspace*{0.7cm}{\cal D}_m W_A = \partial_m  W_A + i \bar{A}_m W_A-i W_A A_m,
\eeq
and similarly for ${\cal D}_m \xi^B$ and ${\cal D}_m \om_B$.
The decomposition of the scalars and fermions into their $SU(2)$ components
has allowed us to split the bosonic as well as the fermionic potential
into $D$--terms and $F$--terms. The precise form of these can be found 
in~\cite{Kristjansen:2008ib}.
The theory has two 't Hooft parameters
\beq
\lambda=\frac{4\pi N}{k},\hspace{0.7cm} 
\hat{\lambda}=\frac{4 \pi M}{k},
\eeq
and one can  consider the double 't Hooft limit
\beq
N,\,M\rightarrow \infty, \hspace{0.7cm} k\rightarrow \infty,
\hspace{0.7cm} \lambda,\, \hat{\lambda} \,\, \mbox{ fixed.}
\eeq
Furthermore, the theory has a multiple expansion in $\lambda$, 
$\hat{\lambda}$, $\frac{1}{N}$ and $\frac{1}{M}$. The action of 
three-dimensional parity flips the levels of the Chern--Simons terms, 
which produces a different theory if $M\neq N$. Thus the ABJ model is 
not parity invariant.

In this letter we will be interested in studying non-planar corrections
(i.e.~$\frac{1}{N}$ and $\frac{1}{M}$ corrections)
for anomalous dimensions at the leading two-loop level.
We shall restrict ourselves to considering scalar
 operators belonging to
a $SU(2)\times SU(2)$ sub-sector i.e.\ operators of the following type
\beq
\label{operators}
{\cal O}= \Tr\left(Z^{A_1} W_{B_1} \ldots Z^{A_L} W_{B_L} \right),
\eeq
where $A_i,B_i \in \{1,2\}$,
and their multi-trace generalizations. A central object in our analysis
will be the parity operation which acts on an operator by inverting the
order of the fields inside each of its traces, i.e.\footnote{We notice
that it is not possible to define in a natural and simple way
a parity operation which acts only on $Z$ or $W$ fields.} 
\beq
P:\hspace{0.3cm} \Tr\left(Z^{A_1} W_{B_1} \ldots Z^{A_L} W_{B_L} \right)
\longrightarrow \Tr\left(W_{B_L} Z^{A_L}\ldots W_{B_1} Z^{A_1} \right).
\eeq
Strictly speaking the parity operation (which would be a true symmetry in ABJM theory)
involves in addition a complex conjugation of the fields~\cite{Bak:2008vd} 
but as complex conjugating the fields inside an operator does not change its anomalous 
dimension the present definition suffices for our purposes.

\section{The derivation of the full dilatation generator \label{derivation}}

\begin{figure}[t]
\begin{center}
\includegraphics[height=4.0cm]{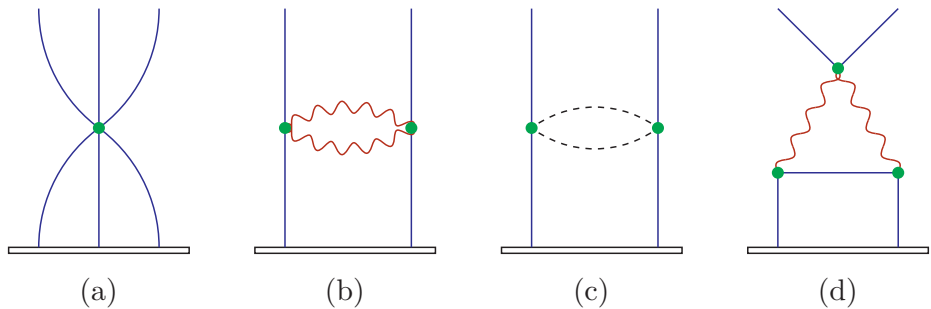}
\end{center}
\caption{The four types of two-loop diagrams contributing to anomalous dimensions. For operators
in the $SU(2)\times SU(2)$ sector diagrams in class (d) do not contribute.}\label{Figure}
\end{figure}

The derivation of the full two-loop dilatation generator of ABJ theory 
is slightly lengthy but follows closely the one for 
ABJM theory~\cite{Kristjansen:2008ib}. 
The contractions one has to do are the same as before, only now one has to
carefully keep track of whether a given contraction gives a factor of $N$ or
a factor of $M$.
The Feynman 
diagrams which contribute at two-loop order consist of the ones depicted
in figure~1 plus 14 self-energy diagrams. All diagrams of course come in
planar as well as non-planar versions. In order to handle most easily the
combinatorics of planar as well as non-planar diagrams it is again convenient
to make use of the method of effective vertices~\cite{Beisert:2002bb}.
An effective vertex is a space-time 
independent vertex which, when contracted with a given operator of the 
type~\rf{operators} gives the combinatorial factor associated with a particular
Feynman integral times the value of the integral. 
If things
work as in ${\cal N}=4$ SYM and as in ABJM theory~\cite{Kristjansen:2008ib}
the contribution from the  bosonic $D$--terms  should cancel against
contributions from gluon exchange, fermion exchange and self-interactions to
all orders in the genus expansion and this is indeed what happens.
To prove this we first calculate the effective vertices corresponding to the
four diagrams in figure~1. We notice, however, that for operators
belonging to the $SU(2)\times SU(2)$ sector there are no contributions
from Fig.~1d. Adding the contributions from the bosonic potential, gluon
exchange and fermion exchange we find
\bea
\lefteqn{\hspace*{-1.0cm}
(V^{bos})^{eff}+(V^{ferm})^{eff}+(V^{gluon})^{eff}}\nonumber \\
&=&
 (V_F^{bos})^{eff}+ V +const
 :\left\{\Tr\left( Z_C^\dg Z^C\right)+
\Tr\left(W_C W^{\dg C}\right)\right\}:, 
\label{Veff}
\eea
where
\beq
const=-\frac{1}{8}(\lambda^2+\hat{\lambda}^2)-
\frac{1}{2}\lambda \hat{\lambda} +\frac{5}{24} 
\frac{\lambda^2}{N^2}+\frac{5}{24}\frac{\hat{\lambda}^2}{M^2}
+\frac{1}{3} \frac{\lambda}{N} \frac{\hat{\lambda}}{M}, 
\label{constant}
\eeq
\vspace*{0.1cm}

\noindent
and where $:\,\,:$ means that self-contractions should be excluded.
The quantity $V$ is a vertex which can be shown to give a vanishing 
contribution when applied to any operator in the $SU(2)\times SU(2)$ sector.
Furthermore, the last term in eqn.~\rf{Veff} has exactly the form expected for 
self-energies  and one can show that it precisely cancels
the contribution from these. To do so one has to check the cancellation
of both the planar and the non-planar part of the constant appearing
in eqn.~\rf{constant}. The planar part of the analysis can be carried
out with the aid of  
reference~\cite{Bak:2008vd}. The non-planar part, however, requires a careful
analysis of the non-planar versions of the 14 self-energy diagrams. 

 Collecting everything, we thus verify
that the full two-loop dilatation generator is indeed given only by the 
$F$--terms in the bosonic potential, i.e.
\bea
D= (V_F^{bos})^{eff}&=&  -\frac{\lambda}{N} \frac{\hat{\lambda}}{M} \,\,
:\Tr \left[ W^{\dg A} Z_B^\dg W^{\dg C} W_A Z^B W_C
-W^{\dg A} Z_B^\dg W^{\dg C}W_C Z^B W_A \right. \nonumber \\
& & \left.
+Z_A^\dg W^{\dg B} Z_C^\dg Z^A W_B Z^C-Z_A^\dg W^{\dg B} Z_C^\dg Z^ C W_B Z^A
\right]:. \label{dilatation}
\eea
It is easy to see that the dilatation generator vanishes when acting
on an operator
consisting of only two of the four fields from the $SU(2)\times SU(2)$ sector.
Accordingly we will denote two of the fields, say
$Z_1$ and $W_1$, as
background fields and $Z_2$ and $W_2$ as excitations. It is likewise easy
to see that operators with only one type of excitation, say $W_2$'s, form
a closed set under dilatations. For operators with only
$W_2\:$-excitations the dilatation generator consists of four terms whereas
in the case with two different types of excitations it has 16 terms.
In both cases $D$ is easily seen to reduce to the
one
of~\cite{Minahan:2008hf,Bak:2008cp}
in the
planar limit
\beq
D_{planar}\equiv \lambda\, \hat{\lambda} D_0= \lambda\,\hat{\lambda}
\sum_{k=1}^{2L} (1-P_{k,k+2}),
\eeq
where $P_{k,k+2}$ denotes the permutation between sites $k$ and $k+2$ and
$2L$ denotes the total number of fields inside an operator. It differs
from the planar dilatation generator of ABJM theory only by having the
pre-factor $\lambda \hat{\lambda}$ instead of $\lambda^2$.
As explained in~\cite{Minahan:2008hf,Bak:2008cp} this is the Hamiltonian
of two alternating $SU(2)$ Heisenberg spin chains, coupled via a momentum 
condition. As mentioned earlier, integrability implies that there exists a 
tower of charges which
all commute and which commute with the Hamiltonian. In particular, there
exists one such charge $Q_3$ which anti-commutes with parity. In addition, the
planar dilatation generator itself commutes with parity, i.e.
\beq
[D_{planar},Q_3]=[D_{planar},P]=\{Q_3,P\}=0.
\eeq
As a consequence, the spectrum of the planar theory has degenerate parity 
pairs, i.e. pairs of operators with identical anomalous dimension but
opposite parity. In reference~\cite{Kristjansen:2008ib} it was shown that
for ABJM theory at the non-planar level the two-loop dilatation generator
still commutes with parity but the degeneracies between parity pairs are
lifted. This hinted towards the absence of higher conserved charges, at least
in a standard form. Below we will analyse the situation for ABJ
theory and find that again the planar degeneracies disappear but in addition
the non-planar two-loop dilatation generator does {\it not} any longer commute with 
parity. 

When acting with the dilatation generator on
a given operator we have to perform three contractions as dictated by the
three hermitian conjugate fields. It is easy to see that by acting with
the dilatation generator one can change the number of traces in a given
operator by at most two.
More precisely, the two-loop dilatation generator has
the expansion
\beq
D= \lambda \hat{\lambda}
\left\{ D_0+\frac{1}{\cal M}\left( D_+ + D_-\right) 
+\frac{1}{{\cal{M}}^2}\left(D_{00}+ D_{++}+ D_{--}\right)\right\}.
\label{Dexpansion}
\eeq
Here $D_+$ and $D_{++}$ increase the number of traces by one and
two respectively
and $D_-$ and $D_{--}$ decrease the number of traces by one and two.
 Finally, $D_0$ does not change the number of traces and 
$D_{00}$ first adds one trace and subsequently removes one or vice versa. 
The quantity $\frac{1}{\cal M}$ stands for $\frac{1}{N}$ or $\frac{1}{M}$ and
$\frac{1}{{\cal M}^2}$ stands for $\frac{1}{N^2}$, $\frac{1}{M^2}$ or
$\frac{1}{MN}$.

Even for short operators it is in practice hard to diagonalise the full
dilatation generator exactly. But one can relatively 
easily diagonalise the planar dilatation generator, either by brute force
or by means of the Bethe equations. Subsequently
the non-planar terms can be treated as perturbations and the energy corrections
found approximately using quantum mechanical perturbation 
theory~\cite{Beisert:2002ff}.
Notice that while energy corrections are generically
of order $\frac{1}{{\cal M}^2}$, degeneracies in the planar spectrum will
lead to energy corrections of order $\frac{1}{\cal M}$. (For details
see~\cite{Kristjansen:2008ib}.)

\section{Short Operators \label{Results}}

In this section we determine non-planar corrections to the anomalous dimensions
of
a number of short operators. This is done by explicitly computing 
and diagonalising the planar
mixing matrix (aided by {\tt GPL Maxima} as well as {\tt Mathematica})
and subsequently determining the non-planar corrections by quantum
mechanical perturbation theory.

\subsection{Operators with excitations on the same chain}

In this sector, the simplest set of operators for which one observes degenerate parity
pairs as well as non-trivial mixing between operators with one, two and
three traces
consists of operators of length 14 with three excitations. There are
in total 17 such non-protected operators. 
Among the
non-protected operators there are only eight which are not descendants.
Their explicit form can be found in reference~\cite{Kristjansen:2008ib}.
The planar anomalous dimensions (in units of $\lambda \, 
\hat{\lambda}$), trace structure and parity for these eight operators,
denoted as ${\cal O}_1,\ldots, {\cal O}_8$, are
\begin{center}
\begin{tabular}{cccc}
Eigenvector & Eigenvalue & Trace structure & Parity\\ \hline
$\CO_1$ &  $5$ & (14) & $-$\\
$\CO_2$ & $6 $ & (2)(12) & $-$\\
$\CO_3$ & $5$ & (14) & $+$  \\
$\CO_4$ &$5+\sqrt{5}$& (2)(12)  & $+$\\
$\CO_5$ & $ 5-\sqrt{5}$ & (2)(12) & $+$\\
$\CO_{6}$& $4$ & (4)(10) & $+$\\
$\CO_{7}$ &$4$ & (2)(2)(10) & $+$\\
$\CO_{8}$ & $6$ & (2)(4)(8) & $+$\\ 
\end{tabular}
\end{center}
We have one pair of degenerate single
trace operators with opposite parity, namely the operators ${\cal
O}_1$ and ${\cal O}_3$.\footnote{We also observe a degeneracy between
the negative parity
double trace state ${\cal O}_2$ and the positive parity
triple trace state $\CO_8$ as well as a degeneracy between the double
trace state $\CO_6$ and the triple trace state $\CO_7$ both of positive
parity.
However, states with different numbers of traces cannot be connected
via the conserved charge $Q_3$.}

Expressing the dilatation generator in the
basis above and taking into account all non-planar corrections we
get (in units of $\lambda \hat{\lambda}$)\footnote{Notice that by construction
the mixing matrix is not hermitian but related to its hermitian conjugate
by a similarity transformation~\cite{Janik:2002bd,Beisert:2002ff}.}
\footnotesize
\[\left(
\begin{array}{cc|cccccc}
5\!+\!\frac{15}{MN}&0&0&
 0&0&0&0 & 0\\
\frac{3}{N}\!+\!\frac{3}{M}&
 \!\!6\!+\!\frac{24}{MN}&0&0&0
& 0 & 0 & 0\\
&&&&&&&
\vspace*{-0.2cm} \\
\hline
\vspace*{-0.2cm}
&&&&&&&\\
0 & 0 &
\,\!\,\!5\!+\!\frac{35}{MN}&0&0& 
-\frac{4}{N}\!-\!\frac{4}{M}&-\frac{4}{MN}&-\frac{2}{MN}\\
0& 0& -\frac{\sqrt{5}/2}{M}\!-\!\frac{\sqrt{5}/2}{N}&\!\!\sqrt{5}\!+\!5\!+
\!\frac{\left(5\!
\sqrt{5}\!+\!35\right)}{MN}&\frac{3\!\sqrt{5}\!-\!5}{MN}
&\!\!\frac{1}{MN}
&0&\!\!\frac{1}{M}\!+\!\frac{1}{N}\\
0& 0& -\frac{\sqrt{5}/2}{M}\!-\!\frac{\sqrt{5}/2}{N}&-\frac{5\!+\!3\!\sqrt{5}}{MN}&
\!\!{5}\!-\!\sqrt{5}\!-\!\frac{5\!\sqrt{5}\!-\!35}{MN}&
-\frac{1}{MN}&\!
\!\,\,0&\!\! -\frac{1}{M}\!-\!\frac{1}{N}\\
0& 0 &-\frac{10}{N}\!-\!\frac{10}{M} &\!\!\frac{4\sqrt{5}+20}{MN} &
-\frac{20-4\sqrt{5}}{MN}&\!\!4\!+\!\frac{28}{MN}&
\,0&0\\
0& 0& 
-\frac{10}{MN}&\frac{2\sqrt{5}+10}{N}\!+\!\frac{2\sqrt{5}+10}{M}&
\frac{2\sqrt{5}-10}{N}\!+\!\frac{2\sqrt{5}-10}{M}&0 &
 \!\!4\!+\!\frac{32}{MN}&-\frac{2}{MN}\\
0& 0& 
-\frac{10}{MN}&\!
\!\frac{12\sqrt{5}+20}{N}\!+\!\frac{12\sqrt{5}+20}{M}&\!
\!\frac{12\sqrt{5}-20}{N}\!+\!\frac{12\sqrt{5}-20}{M}&
\frac{4}{N}\!+\!\frac{4}{M}&-\frac{8}{MN}
&\!6\!+\!\frac{40}{MN}\,\\ 
\end{array} \right).
\]
\normalsize
This mixing matrix of course reduces to that of ABJM theory for $N=M$ as
it should, cf.~\cite{Kristjansen:2008ib}.
We notice that for this  type of operators the positive and negative
parity states still decouple, i.e.\ parity is preserved.
The states ${\cal O}_1$ and ${\cal O}_2$ are exact eigenstates
of the full dilatation generator with non-planar corrections
equal to
\beq
\delta E_1 =  \frac{15}{N M}, \hspace{0.7cm}
\delta E_2 =  \frac{24}{N M}. \hspace{0.7cm}
\eeq
For the remaining operators we observe that all matrix elements between 
degenerate states vanish.
Thus the leading non-planar corrections to the anomalous dimensions
can be found using second order non-degenerate perturbation theory.
The results read 
\bea \delta E_3 &= & \frac{40}{M^2}+\frac{40}{N^2}+\frac{115}{MN}, \hspace{0.7cm}
\delta E_4 = 4(5+2\sqrt{5})\left(\frac{1}{N^2}+\frac{1}{M^2}\right)
+\frac{3(25+7\sqrt{5})}{MN},
\nonumber \\
\delta E_5 &=& 
4(5-2\sqrt{5})\left(\frac{1}{N^2}+\frac{1}{M^2}\right)+
\frac{3(25-7\sqrt{5})}{MN},
\hspace{0.7cm}
\delta E_6  =  -\frac{40}{N^2}-\frac{40}{M^2}-\frac{52}{MN},
\nonumber\\
 \delta E_7  &=& \frac{32}{MN},\hspace{0.4cm} \delta E_8 = 
-40\left(\frac1{N^2}+\frac1{M^2}\right)-\frac{40}{MN}
\eea
We observe that all degeneracies found at the planar level get
lifted when non-planar corrections are taken into account, for
all values of $M$ and $N$.
 This in
particular holds for the degeneracies between the members of the
planar parity pair $({\cal O}_1,{\cal O}_3)$. We have considered
a number of different types of states with only one type of excitation
and have found that the same pattern persists in all cases. In fact,
one can explicitly
show that the matrix elements between $n$ and $(n+1)$--trace
states of the normal ordered operator in
eqn.~\rf{dilatation}, (i.e.\ $D$ without its pre-factor)
can only
depend on $M$ and $N$ through the combination $M+N$. 
Thus one cannot have parity breaking.

\subsection{Operators with excitations on both chains}

The simplest multiplet of operators which have non-planar energy corrections
are operators of length six with two excitations. There are in total three
such non-protected highest weight states. These read
\beq
\begin{split}
{\cal O}_{ 1 }=&\mbox{Tr}(Z_1 W_1 Z_1 W_2 Z_2 W_1)+
\mbox{Tr}(Z_1 W_1 Z_1 W_1 Z_2 W_2 )- 2\mbox{Tr}(Z_1 W_1 Z_2 W_1 Z_1 W_2),\\
{\cal O}_{ 2 }=& \mbox{Tr}(Z_1 W_1 Z_1 W_2 Z_2 W_1)-
\mbox{Tr}(Z_1 W_1 Z_1 W_1 Z_2 W_2), \\
{\cal O}_3 =& \mbox{Tr}(Z_1 W_1) \mbox{Tr}(Z_1 W_1 Z_2 W_2)-
\mbox{Tr}(Z_1 W_1) \mbox{Tr}(Z_1 W_2 Z_2 W_1).\\ 
\end{split}
\eeq
Their associated planar anomalous dimension (in units of 
$\lambda \hat{\lambda}$), parity and trace structure are
\begin{center}
\begin{tabular}{cccc}
Eigenvector & Eigenvalue & Trace Structure &Parity\\ \hline
${\cal O}_1$ & 6  & (6)&$+$\\
${\cal O}_2$ & 6 & (6)&$-$\\
${\cal O}_3$ &  8 & (2)(4)&$-$\\
\end{tabular}
\end{center}
Already in this simple case we have one pair of degenerate states with
opposite parity, namely ${\cal O}_1$ and ${\cal O}_2$.
Expressing the dilatation generator in this basis and taking into account
all non-planar corrections we get (in units of $\lambda \hat{\lambda}$)
\[
\left(
\begin{array}{c|cc}
6 & 0 & \!\frac{1}{M}\!-\!\frac{1}{N}\\
&& \vspace*{-0.35cm} \\ 
\hline 
\vspace*{-0.35cm} && \\
0 &\!6\!-\!\frac{12}{M N}&\!-\frac{3}{M}\!-\!\frac{3}{N} \\
\!\frac{6}{M}\!-\!\frac{6}{N}& \!-\frac{6}{M}\!-\!\frac{6}{N} & 
\!8\!-\!\frac{8}{M N} \\
\end{array} \right).
\]
We observe that in this case the dilatation generator does mix states with
different parity. In other words, the non-planar dilatation generator does
{\it not} commute with $P$. Calculating the energies by second order quantum
mechanical perturbation theory we find
\beq
\delta E_1= -\frac{3}{N^2}-\frac{3}{M^2}+\frac{6}{MN}
,\hspace{0.7cm} \delta E_2 = -\frac{9}{M^2}-\frac{9}{N^2}-\frac{30}{MN}, 
\hspace{0.7cm} 
\delta E_3=\frac{4}{M^2}+\frac{4}{N^2}+\frac{4}{MN}.
\eeq
In particular, we see that the planar degeneracy is lifted.

Let us analyse a slightly larger multiplet of operators with two excitations
of different types that exhibit some more
of the above mentioned non-trivial features of the topological expansion: 
Operators of length eight with
one excitation of each type. There are in total 7 such non-protected
operators. Their explicit form can be found in 
reference~\cite{Kristjansen:2008ib}
and the planar anomalous dimensions (in units of
$\lambda \hat{\lambda} $), trace structure and parity of these operators,
denoted as ${\cal O}_1,\ldots,{\cal O}_7,$ are
\begin{center}
\begin{tabular}{cccc}
Eigenvector & Eigenvalue & Trace Structure &Parity\\ \hline
 $\CO_1$ & 8  & (8)&$-$\\
$\CO_2$ & 4 & (8)&$-$\\
 $\CO_3$ &  8 & (4)(4)&$-$\\
$\CO_4$ & 6 & (2)(6)&$-$\\
$\CO_5$ & 8 & (2)(2)(4)&$-$  \\
$\CO_6$ & 4& (8)&$+$\\
$\CO_7$ & 6 & (2)(6)&$+$\\
\end{tabular}
\end{center}
Notice that we have two pairs of degenerate operators with
opposite parity, namely the single trace operators $\CO_2$ and $\CO_6$ and
the double trace operators $\CO_4$ and $\CO_7$.\footnote{The double trace
operators ${\cal O}_4$ and ${\cal O}_7$
can be related via $Q_3$ when letting $Q_3$ act only on the
longer of the two constituent traces of the operators.}

Expressing the dilatation generator in the basis given above and taking into
account all non-planar corrections we get 
(in units of $\lambda \hat{\lambda}$)
\[ \left(
\begin{array}{ccccc|cc}
8&\frac{8}{\:M N}&\!\frac{8}{N}\!+\!\frac{8}{M}&\!\frac{2}{N}\!+\!\frac{2}{M}&-\frac{8}{\:M N}&0&\!\frac{2}{M}\!-\!\frac{2}{N}\cr 
\frac{8}{\:M N}&\!4\!-\!\frac{12}{M N}&0&-\frac{1}{N}\!-\!\frac{1}{M}&-\frac{4}{\:M N}&0&\!\frac{1}{N}\!-\!\frac{1}{M} \cr 
\!\frac{8}{N}\!+\!\frac{8}{M}&\!-\!\frac{4}{N}\!-\!\frac{4}{M}&8 &0&0&\!\frac{4}{M}\!-\!\frac{4}{N}&0\cr 
0&\!-\!\frac{8}{N}\!-\!\frac{8}{M}&-\frac{8}{\:M N}&\!6\!-\!\frac{8}{\:MN}&\!-\!\frac{6}{N}\!-\!\frac{6}{M}&\!\frac{4}{M}\!-\!\frac{4}{N}&0 \cr 
0&\frac{8}{\:M N}&0&\!-\!\frac{6}{N}\!-\!\frac{6}{M}&\!8\!-\!\frac{8}{\:MN}&0&\frac{6}{N}\!-\!\frac{6}{M}\cr
&&&&&& \vspace*{-0.3cm} \\
\hline
\vspace*{-0.3cm} 
&&&&&& \\ 
0&0&0&\frac{1}{M}\!-\!\frac{1}{N}&0&\!4\!+\!\frac{4}{MN}&\frac{1}{N}\!+\!\frac{1}{M}\cr 
0&0&0&0&\!\frac{2}{N}\!-\!\frac{2}{M}&\!\frac{4}{N}\!+\!\frac{4}{M}&\!6\!+\!\frac{8}{\:MN}\cr 
\end{array}\right).
\]
This mixing matrix of course reduces to that of ABJM theory for $N=M$ as
it should, cf.~\cite{Kristjansen:2008ib}.
We observe again that  the dilatation generator does mix states with
different parity.
To find the corrections to the eigenvalues we use perturbation theory 
as described in section~\ref{derivation}.
First, we notice that most matrix elements between degenerate states vanish.
The only exception are the matrix elements between the
states $\CO_1$ and $\CO_3$. To find the non-planar correction to the energy
of these states we diagonalise the Hamiltonian in the corresponding subspace
and find 
\beq\label{degstates} \delta E_{1,3}=\mp \left( \, \frac{8}{N}+\frac{8}{M}\,\right). \eeq
For the remaining operators the leading non-planar corrections to
the energy can be found using second order non-degenerate
perturbation theory. The results read 
\beq  
\delta E_2 =  -\frac{20}{N M}-\frac{4}{N^2}-\frac{4}{M^2}, 
\hspace{0.4cm}  \nonumber
\delta E_4 =  -\frac{40}{N M}-\frac{12}{N^2}-\frac{12}{M^2}, 
\eeq
\beq
\delta E_5 = \frac{16}{N M}+\frac{24}{N^2}+\frac{24}{M^2},\hspace{0.4cm} 
\delta E_6 = \frac{4}{MN} -\frac{4}{N^2}-\frac{4}{M^2},
\hspace{0.4cm} 
\delta E_7 = \frac{24}{MN}-\frac{4}{N^2}-\frac{4}{M^2}. 
\nonumber 
\eeq
We again notice that all
degeneracies observed at the planar level get lifted when non-planar
corrections are taken into account, for all values
of $M$ and $N$. This in particular holds for the
degeneracies between the members of the two parity pairs.
We have examined a number of operators with excitations of two different
types and found that the same pattern persists in all cases.  A closer 
scrutiny of the action of the dilatation generator reveals that the asymmetry
between $M$ and $N$ originates from the situation where the 
operator separates two neighbouring excitations, a situation which one does 
not encounter when the two excitations are on the same chain. 
Let us note that the characteristic polynomial of the anomalous dimension matrices will 
always be even in $M-N$. This implies that the eigenvalues will generically be even under 
the interchange of $M$ and $N$ (as is the case above). A possible exception might arise in
cases where nonzero matrix elements appear between planar degenerate states which have opposite parity
and differ in trace number by one (notice that the requirement of different trace structure prevents
this complication from arising for planar parity pairs). Although mixing of the above type does
occur, we did not observe any asymmetry in the eigenvalues for the explicit cases we examined.

\section{Conclusion \label{conclusion}}

We have derived and analysed the non-planar corrections to the two-loop dilatation 
generator of ABJ theory in the $SU(2)\times SU(2)$ sub-sector.
Our analysis shows that these corrections mix states with positive and negative parity, i.e.
\beq
[D_{non-planar}^{ABJ},P] \neq 0.
\eeq
More precisely, the value of the commutator is proportional to $M-N$. 
This is in contrast to earlier studies of the \emph{planar} two-loop dilatation 
generator which did not reveal any sign of parity 
breaking~\cite{Bak:2008vd,Minahan:2009te}. Furthermore, whereas the planar dilatation
generator could be proved to be integrable, we do not see any indication of this
being the case for the non-planar one, since none of the planar degeneracies between 
parity pairs survive the inclusion of non-planar corrections. It is an interesting 
question whether the planar dilatation generator remains integrable and
parity invariant when higher loop corrections are taken into account. In
this connection it is worth mentioning that parity breaking does not
prevent integrability~\cite{Bak:2008vd,Minahan:2009te}. At planar level, one could try 
to address the question of parity breaking at higher-loop order from the 
string theory side by calculating a transition amplitude between
two string states of different parity living in an instanton background of
the ABJ theory dual. We note that an interesting effect of parity breaking in the 
non-interacting string theory has been observed in~\cite{Bergman:2009zh}.

One could also try to match the results of the present
calculation to the behaviour of the dual string theory by  
calculating the semi-classical amplitude for non-parity-conserving
splitting of a one-string state into a two-string state in the spirit
of~\cite{Peeters:2004pt,Casteill:2007td}. Of course, this calculation would at best allow 
us to obtain qualitative agreement between non-planar gauge theory and interacting string theory. 
How to achieve quantitative agreement remains a challenge.

\vspace*{0.5cm}

\noindent
{\bf Acknowledgments:}
We thank N.\ Beisert, T.\ Harmark, M.\ Orselli,
A.\  Wereszczynski, K.\ Zarembo and especially S.\ Hirano for 
useful discussions. 
CK and KZ were supported by FNU through grant number 272-06-0434. 
PC was supported in part by the Niels Bohr International Academy.

\providecommand{\href}[2]{#2}\begingroup\raggedright\endgroup

\end{document}